# Loss free shaping of few-cycle terawatt laser pulses

L.M. Railing[1], M.S. Le[1], C.M. Lazzarini[2,3], and H.M. Milchberg[1,4*]

[1] *Institute for Research in Electronics and Applied Physics and Dept. of Physics, University of Maryland, College Park, MD 20742, USA*
[2] *ELI Beamlines Facility, The Extreme Light Infrastructure ERIC, Dolni Brezany, 25241, Czech Republic*
[3] *Faculty of Nuclear Sciences and Physical Engineering, Czech Technical University, Prague, 11519, Czech Republic.*
[4] *Dept. of Electrical and Computer Engineering, Univ. of Maryland, College Park, MD 20742, USA*
*Corresponding author: milch@umd.edu*

**We demonstrate loss-free generation of 3 mJ, 1 kHz, few-cycle (5 fs at 750 nm central wavelength) double pulses with separation from 10 fs to 100 fs, using a helium filled hollow core fiber (HCF) and chirped mirror compressor. Crucial to our scheme are simulation-based modifications to the spectral phase and amplitude of the oscillator seed pulse to eliminate the deleterious effects of self-focusing and nonlinear phase pickup in the chirped pulse amplifier. The shortest pulse separations are enabled by spectral reshaping and pulse splitting in the HCF compressor.**

Fine control of the temporal shape of ultrashort pulses produced by chirped pulse amplification (CPA) systems [1] is desirable for many of their applications, including high harmonic generation [2,3], attosecond interactions and probing of electronic systems [4], and laser wakefield acceleration [5].

Traditional shaping of high-power ultrashort pulses has been accomplished by passing pre- or post-amplification pulses through dispersive optics and a spatial light modulator (SLM) [6], or by passing pre-amplification seed pulses through an acousto-optic programmable dispersive filter (AOPDF) [7]. While the pulse shaping device itself is often inefficient, lost energy can be regained in a regenerative amplifier (RGA)-based CPA system with additional round trips in the RGA. Any undesired spectral phase accumulated in the additional round trips can be pre-compensated by the SLM or AOPDF [8]. However, as CPA systems have become more energetic, a potential problem is presented by nonlinear effects during amplification and potential damage to the RGA. If the ultimate aim is few-cycle pulses, nonlinear compression techniques based on self-phase modulation (SPM) [9,10,11] can be complicated by nonlinear effects in the RGA. Because of these difficulties, prior pulse shaping efforts have sacrificed either a large fraction of the pulse energy [12,13] or some versatility of the pulse shaper [14].

One of the most difficult pulse shapes to produce in the few-cycle regime, without loss of energy, is the double pulse: two few-cycle pulses separated by an adjustable time delay. Such pulses have been used in, for example, pump-probe measurements of electronic states in atoms [4] and molecules [15], as well as in the production of spectrally tunable attosecond pulses [16]. Single few-cycle pulse generation is now routinely achieved using gas-filled hollow core fiber (HCF)-based pulse compressors [9]. One method for generating double few-cycle pulses is phase front splitting and delay with segmented mirrors [16]; here beam quality may suffer from diffraction. Another method uses spectral amplitude and phase shaping after the HCF but before the compressor; this can result in pulse energy loss of up to 80% [15, 17]. Mach-Zehnder geometry can also be used, but half the initial pulse energy is discarded at the second beam splitter. The use of birefringent calcite plates for pulse division at the HCF entrance can mitigate such energy loss [18]; however, the price paid is that the two pulses are orthogonally polarized and their temporal separation is fixed, aspects which may not be suitable for some applications. In general, splitting CPA pulses prior to injection into an HCF can result in reduced coupling into the HCF from phase front distortions [19] and linear losses from the additional optics. A more robust and efficient shaping technique is required to handle all of these challenges.

In this paper, we present a method for shaping of terawatt-scale few-cycle double pulses by control of the pre-amplification spectral amplitude and phase of the RGA seed pulse. Self-focusing damage in the RGA is avoided and nonlinear phase effects are compensated, while post-amplification nonlinear effects in a helium-filled HCF are exploited for generation of the desired double pulses. We demonstrate generation of 3 mJ, co-propagating and co-polarized few-cycle double pulses with a delay tunable over $\sim 10 - 100$ fs with no energy loss.

Our technique uses the optical setup shown in Fig. 1, already in use for our experiments in few-cycle laser-driven wakefield acceleration [20]. Pulses from a mode locked Ti: Sapphire oscillator are stretched to ~180 ps in the grating stretcher of a 1 kHz pulse repetition rate CPA system, and then shaped in spectral amplitude and phase using an AOPDF before amplification in a Ti: Sapphire RGA followed by a single pass amplifier (SPA). The pulse is then compressed in a grating compressor (the bandwidth-limited single pulse width is $\tau_0 \sim 35$ fs), injected into and nonlinearly broadened in a 2.5-m-long helium filled HCF, and then compressed in a chirped mirror (CM) compressor. A subsequent prism pair and windows of varying thickness are used to tune dispersion to produce the desired pulse shapes. The total pulse energy is ~3 mJ regardless of the pulse shape. We temporally characterize the CPA output pulses with a single-shot SHG FROG [21] before injection into the HCF. After the CM compressor, the final pulse shape is measured by a separate broadband, single shot SHG FROG [22].

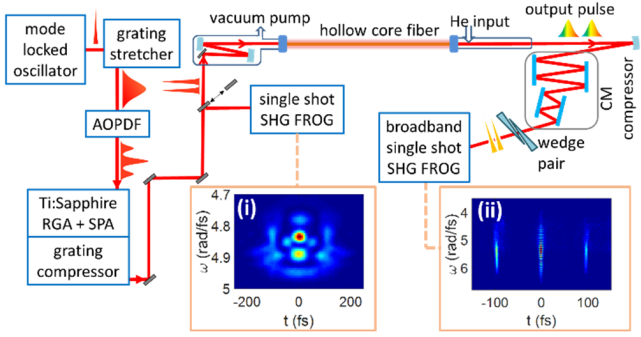

**Fig. 1.** Seed laser pulses are stretched and shaped in spectral amplitude and phase with an AOPDF before amplification in regenerative amplifier (RGA) and single pass amplifier (SPA). After compression, the pulses are broadened in a differentially pumped hollow core fiber (HCF) filled with helium. After broadening, a chirped mirror (CM) compressor, a set of windows of varying thickness, and a prism pair apply tunable negative dispersion to the pulse to reach optimal compression. The pulse is characterized using second harmonic generation frequency resolved optical gating (SHG FROG) devices both before the HCF entrance and after the CM compressor. Typical SHG FROG traces (i) before the HCF entrance, and (ii) after the CM compressor.

The first step in generating few-cycle double pulses requires AOPDF modification of the stretched pulse to give the complex spectral amplitude $\tilde{E}(\omega) = \alpha \tilde{E}_0(\omega) \cos((\omega - \omega_0)\Delta t/2)$, where $\tilde{E}_0(\omega)$ is the original spectral amplitude, $\omega_0$ is the central frequency, $(\Delta\omega)_0$ is the FWHM bandwidth, $\Delta t_{AOPDF} \equiv \Delta t$ is the nominal pulse separation set by the AOPDF, and $\alpha$ is a coefficient accounting for losses. Because of the extremely large quadratic spectral phase imposed by the stretcher ($\sim 10^6$ fs$^2$), each narrow spectral slice is associated with a narrow temporal slice along the pulse. The beating-modulated spectral and temporal intensities of the seed pulse, $I_\omega(\omega) \propto |\tilde{E}(\omega)|^2$ and $I_t(t) \propto |E(t)|^2$, are therefore effectively the same shape upon injection into the RGA. During amplification in the RGA, these intensity modulations can lead to damage in the Ti:Sapphire or Pockels cell KD*P crystals due to Kerr self-focusing.

Potential damage from self-focusing can be mitigated with an AOPDF-applied amplitude dip around the spectral peak at $\omega = \omega_0$, ($\lambda_0 = 800$ nm) so that the RGA input pulse becomes $\tilde{E}_{in}(\omega) = \Gamma(\omega)\tilde{E}(\omega)$, where $\Gamma(\omega) = \left(1 - A e^{-(\omega-\omega_o)^2/\Delta^2}\right)$ is the dip function. Here, the optimal spectral dip and width parameters, $A$ and $\Delta$, are first estimated by pulse propagation simulations in the RGA and then refined by experiment. The simulations propagate the field over multiple passes through the RGA, accounting for gain and thermal lensing in the gain (Ti:Sapphire) crystal and the Kerr effect in the gain crystal and two Pockels cell crystals. Amplification of the pulse amplitude in the gain crystal is modeled by [23]

$$\partial I(r,z,t)/\partial z = \sigma_e(\lambda(t))N(r,z,t)I(r,z,t) \quad (1)$$

$$\partial N(r,z,t)/\partial t = -\sigma_e(\lambda(t))N(r,z,t)\lambda(t)I(r,z,t)(hc)^{-1}, \quad (2)$$

so that the field at the crystal exit, $z = z_c$, after each pass is $E(r,z_c,t) \propto \sqrt{I(r,z_c,t)} \exp[ik \int_0^{z_c} n_2^s I dz + \Delta\phi_{th}(r,z_c)]$. Here, $I(r,z,t)$ is the spatiotemporal intensity, $N$ is the number density of upper state titanium ions in the Ti:Sapphire crystal, $\sigma_e(\lambda)$ is the stimulated emission cross section, $\lambda(t)$ is the instantaneous wavelength vs. time of the strongly chirped pulse, $n_2^s$ is the nonlinear index of sapphire [24], and $\Delta\phi_{th} = -kr^2/2f_{th}$ is the phase profile imposed by thermal lensing, where $f_{th} = 95$ cm is the focal length calculated from the pump profile [25]. The field also picks up the nonlinear phase factor $\exp(ikn_2^p I z_p)$ at each of two quarter-wave Pockels cells, where $n_2^p$ and $z_p$ are the nonlinear index [26] and length of the Pockels cell crystal. The chirped pulse mapping of time to frequency enables numerical propagation of each wavelength slice of the field, via split-step Fourier solution of the Helmholtz wave equation [27], to the cavity end mirrors and back to the gain medium and Pockels cells, tracing out the desired number of round trips in the RGA. The input pulse to the simulation was a 3.4 nJ, 180 ps duration linearly chirped pulse of the form of $\tilde{E}_{in}(\omega)$, with an initial spot size twice larger than the $1/e^2$ intensity cavity mode radius of $\sim$330 μm at the gain rod. The simulation models the gain profile on the RGA crystal as a 4th order super-Gaussian with $\sim$430 μm HWHM, a fit to the pump spot imaged through the cavity end mirror, and the initial upper state population $N$ was set by the measurement of small signal gain. The propagation simulations were run with varying values of $A$ and $\Delta$ in $\Gamma(\omega)$ until intensity spikes were eliminated in the gain rod and Pockels cells. The values $A = 0.5$ and $\Delta = 0.03$ rad/fs were found to work well for all cases in this paper.

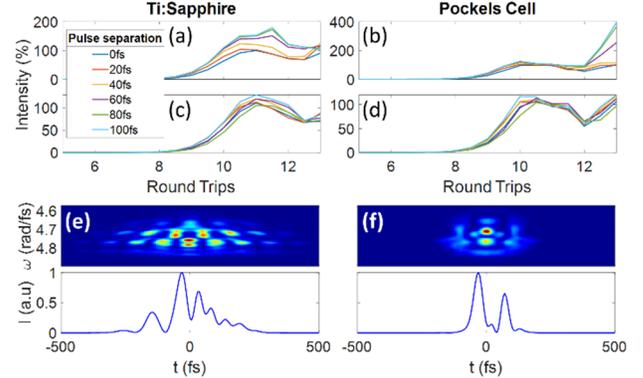

**Fig. 2. (a)** Peak pulse intensity in gain rod (as a percentage of peak intensity of the unmodulated pulse) vs. pulse separation $\Delta t$ and number of RGA round trips without dip function $\Gamma(\omega)$ applied to the injected pulse. **(b)** Same for Pockels cell. **(c)** and **(d)**: peak intensities in gain rod and Pockels cell with dip function applied. **(e)** For $\Delta t = 100$ fs, FROG trace (top) and extracted temporal intensity (bottom) without pre-compensating phase $\Delta\varphi_{pre}(\omega)$ applied at AOPDF. **(f)** Desired double pulse structure achieved when $\Delta\varphi_{pre}(\omega)$ is applied.

Figures 2(a) and (b) plot the simulated peak intensity in the gain rod and Pockels cell vs. round trip *without* the dip function $\Gamma(\omega)$ applied to the injected pulse. The peak intensity, scaled as a percentage of peak intensity of the unmodulated pulse, is seen to grow to maximum levels of $\sim$150% in the gain rod and $\sim$400% in the Pockels cells for $\Delta t = 100$ fs. With the dip function applied, Figs. 2(c) and (d) show peak intensities reduced to $\sim$120% in each element. We experimentally verified the efficacy of the dip function by monitoring the FWHM pulse spectrum in the RGA; it did not narrow noticeably owing to self-focusing.

Figure 2(e) is a FROG trace of the RGA output for $\Delta t = 100$ fs: with many modulations, it is far from the desired double pulse shape. The deleterious effect leading to this distorted pulse is the nonlinear phase pickup, $\Delta\varphi_{NL}(\omega) \propto I_\omega(\omega)$, by the modulated chirped pulse in the RGA during amplification of $\tilde{E}_{in}(\omega)$. This effect was mitigated by applying a pre-compensating phase at the AOPDF, $\Delta\varphi_{pre}(\omega) = -\Delta\varphi_0 I_\omega(\omega)/I_{\omega,max}$, so that the RGA seed pulse becomes $\tilde{E}_{in}(\omega) = \exp(i\Delta\varphi_{pre}(\omega))\,\Gamma(\omega)\tilde{E}(\omega)$. Since $\Delta\varphi_{NL}(\omega)$ is largely accumulated in the last few round trips when gain narrowing has ceased, the measured spectrum after amplification was used for $I_\omega(\omega)$ [28]. It was found that for the 6.5 mJ RGA pulse energies of our experiments, setting $\Delta\varphi_0 \sim 1.75$ rad eliminated the modulations and produced the desired $\Delta t = 100$ fs double pulse structure plotted in Fig. 2(f). This is consistent with our simulations.

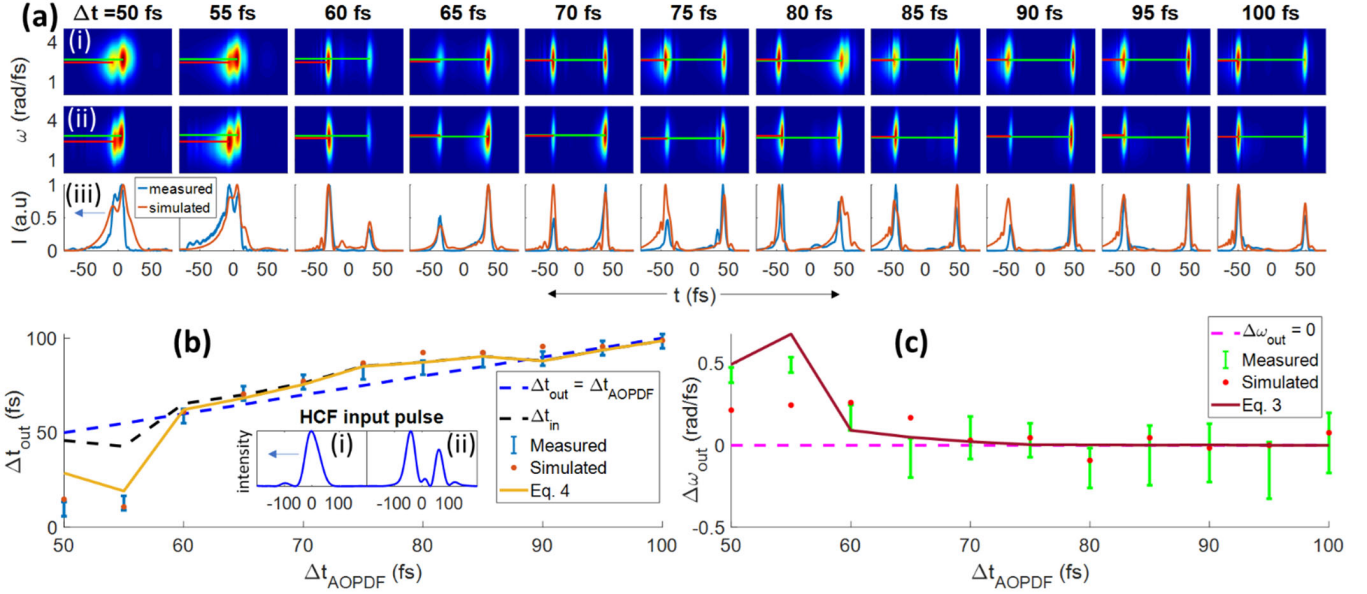

**Fig. 3. (a)** For $\Delta t = 50 - 100$ fs, (*i*) Plot of $|E_{Gs}(\omega,t)|^2$, where $E_{Gs}(\omega,t)$ is the Gabor transform of the simulated field , (*ii*) plot of $|E_{Gm}(\omega,t)|^2$, where $E_{Gm}(\omega,t)$ is the Gabor transform of the measured field retrieved from SHG FROG traces, and (*iii*) plots of corresponding temporal intensities for measured (blue curves) and simulated (orange curves) fields at the exit of the CM compressor. **(b)** CM compressor output pulse separation $\Delta t_{out}$ vs. $\Delta t$: measured (blue points), simulated (orange points), simple model of Eq. 4 (tan line), setting $\Delta t = \Delta t_{in}$ (see text). Inset (i): measured pulse at HCF entrance for $\Delta t = 50$ fs. Inset (ii): same, for $\Delta t = 100$ fs. **(c)** Spectral separation $\Delta\omega_{out}$ between the peaks after the CM compressor vs. $\Delta t$. The horizontal red and green lines in (a)(i) and (a)(ii) mark the central frequencies in each peak used, respectively, for the simulated (red points) and measured (green points) spectral separations. Overlaid is the prediction of Eq. 3 (brown curve), where $\Delta\omega_{out} = 2|\delta\omega(t = \mp\Delta t/2)|$. All error bars are calculated from the imaging resolution of the SHG FROG measurements.

An additional consideration plays a crucial role for $\Delta t < 2\tau_0$ (~70 fs). For that range of $\Delta t$, intra-pulse spectral interference in the RGA becomes sensitive to higher order spectral phase introduced by the pulse stretcher, resulting in distorted output pulse shapes from the CM compressor. This problem is mitigated by using the AOPDF to subtract the high order spectral phase, $\Delta\varphi_{h.o.}(\omega)$, so that the RGA input pulse becomes $\tilde{E}_{in}(\omega) = \exp[i(\Delta\varphi_{pre}(\omega) - \Delta\varphi_{h.o.}(\omega))]\,\Gamma(\omega)\tilde{E}(\omega)$. Since $\Delta\varphi_{h.o.}(\omega)$ is largely introduced by the stretcher, it is determined by FROG measurements of a single pulse at the CPA system exit.

In practice, tuning $\Delta t$ requires adjusting the helium pressure in the HCF to maintain maximum bandwidth. For $\Delta t > 2\tau_0$, the duration of a single pulse, the peak intensity of the double pulse at the HCF entrance is roughly halved. To maintain the SPM bandwidth at the HCF exit, the nonlinear index of helium, $n_2^h$ [29], was doubled by doubling the helium pressure from the single pulse case. For $\Delta t < \sim 2\tau_0$, the helium pressure was increased by ~50-65%.

For each pulse shape generated at the exit of the CM compressor, we recorded the corresponding SHG FROG trace of the pulse at the HCF entrance. To gain insight into the pulse compression physics, the extracted complex field envelope was used as the input pulse to an HCF propagation simulation using the multimode generalized nonlinear Schrödinger equation [30]. The simulation included the first two hybrid transverse electric modes only, based on the visible longitudinal beat period of ~0.27 m period observed for HCF leakage. The intensity weighting of each mode (0.93 and 0.07) at the HCF entrance was based on the throughput at low input pulse energies (78%). The simulated pulse at the CM compressor exit, $E_{sim}(t)$, was obtained by applying the spectral phase from the CM compressor to the simulated HCF exit pulse.

For a range of $\Delta t$ up to 100 fs, Fig. 3(a) plots (i) the magnitude squared of the Gabor transform of $E_{sim}(t)$, (ii) the magnitude squared of the Gabor transform of $E_{meas}(t)$, the complex field measured by SHG FROG at the CM compressor exit, and (iii) the measured pulse intensity $|E_{meas}(t)|^2$ vs. time (blue curves) overlaid with $|E_{sim}(t)|^2$. Agreement between experiment and simulations is good. It is interesting to note that for temporally symmetric pulses injected at the HCF, the compressed HCF output pulses can be asymmetric, with either the leading or trailing pulse more intense than the other. This is caused by the complex effects of self-steepening during propagation in the HCF, and its effects are well-predicted by our simulations.

For $\Delta t > 60$ fs, the separation, $\Delta t_{out}$, of the two peaks in $|E_{meas}(t)|^2$ follows $\Delta t_{out} \sim \Delta t$, as also seen in Fig. 3(b). However, for $\Delta t < \sim 60$ fs, $\Delta t_{out}$ shrinks much faster than $\Delta t$, leading to peak separations $\Delta t_{out} \sim 10\text{-}13$ fs at the compressor exit for $\Delta t \sim 50\text{-}55$ fs at the HCF entrance. For $\Delta t < 50$ fs, pulse shapes at the CM compressor exit are indistinguishable from single peaks. Insets (*i*) and (*ii*) in Fig. 3(b) plot SHG-FROG-measured pulse shapes at the HCF entrance for (*i*) $\Delta t = 50$ fs and (*ii*) $\Delta t = 100$ fs; in general, evolution of pulses with $\Delta t < \sim 60$ fs in the RGA produces merged peaks and a longer pulse rather than separated short peaks, owing mainly to gain narrowing.

This effect motivates a simple model for how $\Delta t < \sim 60$ fs leads to much shorter $\Delta t_{out}$. For successful pre-compensation of nonlinear and high order phase pickup (see earlier discussion), neglecting the dip function, and accounting for gain narrowing, the compressed CPA system output is the Fourier transform of $\tilde{E}_{in}(\omega)$, giving at the HCF entrance the pulse envelope $E(t) = E_0[\exp(-\frac{1}{4}\sigma^2(t+\Delta t/2)^2) + \exp(-\frac{1}{4}\sigma^2(t-\Delta t/2)^2)]$, where the gain-narrowed FWHM bandwidth is $\Delta \omega = 2\sqrt{\ln 2}\ \sigma \sim 0.6(\Delta \omega)_0$. As a result, for $\Delta t < 60$ fs, $|E(t)|^2$ appears as a single widened peak with a nearly flat top. With the nonlinear frequency shift in the HCF approximated as $\delta\omega(t) = -k_0 n_2^h L\ (d/dt)|E(t)|^2$, where $L$ is the HCF length and $k_0$ is the central longitudinal wavenumber of the HCF mode, the main shifts occur at the widened pulse's leading and trailing edges, with little shift contributed by the pulse center region. It is then straightforward to show that the frequency shift at the leading and trailing edges is

$$\delta\omega\left(t = \mp\frac{\Delta t}{2}\right) = \mp \frac{\sigma^3 k_0 n_2^h L U \Delta t}{2\sqrt{2\pi} A_{eff}}\ e^{-5\xi} \cosh(2\xi)\operatorname{sech}(\xi), \quad (3)$$

where $\xi = \sigma^2 \Delta t^2/16$, U is the total energy of the pulse, and $A_{eff}$ is the effective $HE_{11}$ mode area in the HCF.

The CM compressor's negative group dispersion ($\varphi_{CM}^{(2)} \sim -(35-50)\ \text{fs}^2$) delays the leading redshifted portion of the pulse more than the trailing blue shifted portion. This causes pulse splitting and generation of a pair of pulse peaks at the CM compressor exit with separation much smaller than $\Delta t$,

$$\Delta t_{\text{out}} = \Delta t + 2\varphi_{CM}^{(2)}\ |\delta\omega(t = \mp\Delta t/2)|. \quad (4)$$

A similar effect has been seen previously for SPM of super-Gaussian pulses in negatively dispersive fiber [31]. If we use values of $\Delta t$ obtained from fits of $E(t)$ to the SHG FROG measurements at the HCF entrance ($\Delta t_{in}$, see Fig. 3(b)), then Eqs. (3) and (4) mostly agree with the measured and simulated compressed pulses, as seen in Fig. 3(b) and (c), with the simulations underestimating the spectral separation between the two pulses (Fig. 3(c)).

In summary, we have demonstrated loss free shaping of terawatt-scale few-cycle pulses. We use simulation-motivated settings of an acousto-optic programmable dispersive filter to mitigate the effects of self-focusing and nonlinear phase pickup in the high amplification section (here the regenerative amplifier) of a chirped pulse amplification system. We have used this method to produce few-cycle double pulses with tunable delay from 10-100 fs without loss of energy.

This work was supported by the US Dept. of Energy (DESC0024406), the National Science Foundation (PHY2010511), and the Air Force Office of Scientific Research (FA9550-21-1-0405). The authors thank Scott Hancock, Ilia Larkin, and Bo Miao for discussions and technical assistance.